\documentstyle[11pt,twoside,jltp,epsfig]{article}

\runninghead{A.A. Aligia {\it et al.}}{Superconductivity and 
Topological Numbers}
\begin{document}

\title{Superconductivity and Topological Numbers in the Hubbard Chain 
with Bond-Charge Interaction}
\author{A.A. Aligia, K. Hallberg, C.D. Batista 
\address{Centro At\'omico Bariloche and Instituto Balseiro,
Comisi\'on Nacional de Energ{\'\i}a At\'omica,
8400 S.C. de Bariloche, Argentina
} and G. Ortiz%
\address{Theoretical Division, 
Los Alamos National Laboratory, Los Alamos, NM 87545}}

\begin{abstract}
We determine the quantum phase diagram of the Hubbard chain with electron-hole 
symmetric correlated hopping at 1/2- and 1/4-filling using geometric concepts 
and continuum limit field theory. The long distance behavior of various correlation 
functions show a very rich phase diagram 
with several insulating, metallic, and superconducting phases, which might be relevant 
to (TMTSF)$_2$X compounds. The closing of charge and spin gaps are accurately resolved 
as topological transitions (jumps in $\pi$ of Berry phases). The 
metallic or insulating character of each thermodynamic phase is obtained from the 
ground-state expectation value of a displacement operator in reciprocal space.
 
\end{abstract}

\maketitle

\vspace{0.3in}


Superconductivity studies in the Bechgaard salts with the generic formula 
(TMTSF)$_{2}$X have enriched our knowledge of low-dimensional conductors with the 
new physics discovered \cite{jer}. Although at very low temperatures, the 
transverse hopping should play a role, the extremely large anisotropy suggests that 
a one-dimensional (1D) model should be a good starting point for understanding 
these compounds. Tight-binding calculations suggest that only one band crosses the 
Fermi level \cite{gra}. Therefore, one can construct an effective 1D one-band model, 
retaining only the ground state (GS) of each unit cell for zero, one, and two holes 
and can map these states into the corresponding ones of a Hubbard model. This is a 
formidable task, in contrast to 
cuprate superconductors \cite{sim}, and has not been done yet. However,
it is evident that the effective hopping should depend upon the occupation of the 
sites involved, and one expects an important on-site interaction $U$. These physical 
considerations lead to the following effective Hamiltonian
\begin{eqnarray}
&&H=\sum_{\langle i,j\rangle \sigma }(c_{i\sigma }^{\dagger }c_{j\sigma
}^{\;}+h.c.)\{t_{AA}(1-n_{i\bar{\sigma}})(1-n_{j\bar{\sigma}})+t_{BB}\ n_{i%
\bar{\sigma}}n_{j\bar{\sigma}}  \nonumber \\
&&+t_{AB}\left[ n_{i\bar{\sigma}}(1-n_{j\bar{\sigma}})+n_{j\bar{\sigma}%
}(1-n_{i\bar{\sigma}})\right] \}+\ U\sum_{i}(n_{i\uparrow }-\frac{1}{2}%
)(n_{i\downarrow }-\frac{1}{2})\ ,  \label{eq1}
\end{eqnarray}
in standard notation, with one hole per unit cell (1/2-filling, $n=1$). We restrict 
the present study to the electron-hole symmetric case $t_{AA}=t_{BB}=1$.

Another possible starting point is to retain the states of lowest energy of only 
one organic molecule TMTSF (instead of the unit cell). The resulting Hamiltonian 
is similar to (\ref{eq1}), but now the operators $c_{i\sigma }^{\dagger }$, 
$c_{i\sigma}^{\;}$ act on low-energy states of one molecule (labeled by $i$), 
the parameters are different, and the appropriate filling is one hole each two 
sites (1/4-filling, $n=1/2$).

In this work we determine the quantum phase diagram of $H$, both for 1/2- and 
1/4-filling, using recent developments based on geometrical concepts 
\cite{top,res,z}, 
Lanczos diagonalization, and density-matrix renormalization group (DMRG) methods. 
We also extend away from 1/2-filling previous studies using continuum limit 
field theory (CLFT) \cite{jap}, valid for small values of the interactions. 
Varying the parameters of the model, we obtain several thermodynamic phases 
with noticeable correspondence to the observed ones.

The basic tool for determining accurate phase boundaries are the charge 
($\gamma _{c}$) and spin ($\gamma _{s}$) Berry phases. $\gamma _{c}$ ($\gamma _{s}$) 
is the phase captured by the GS of a system of length $L$, when twisted boundary 
conditions $c_{i+L\sigma }^{\dagger}=e^{i \Phi_{\sigma }}c_{i\sigma }^{\dagger }$ 
are varied in the cycle $0\leq \Phi \leq 2\pi $, keeping 
$\Phi =\Phi _{\uparrow }=\Phi _{\downarrow}$ 
($\Phi =-\Phi_{\uparrow }=\Phi _{\downarrow }$) \cite{top}. For systems with 
inversion symmetry, $\gamma _{c}$ and $\gamma _{s}$ can only be either 0 or $\pi$ 
(mod $2\pi$). Then, there are only four possible values of the topological vector 
$\vec{\gamma} =  (\gamma_{c},\gamma _{s})$. In this model, for $n=1$, these values 
correspond to the four thermodynamic phases which can be obtained by  varying the 
parameters. A jump by $\pi$ in $\gamma _{c}$ ($\gamma _{s}$) signals the opening 
of a charge $\Delta_c$ (spin $\Delta_s$) gap and a change by $e/2$ in the 
polarization (difference between up and down polarizations) of the system. For the 
present $H$ and any filling $\gamma _{s}=\pi$ if and only if $\Delta_s$=0, and for 
$n=1$, $\gamma _{c}=0$ if and only if $\Delta_c$=0. 

\begin{figure}
\centerline{\psfig{file=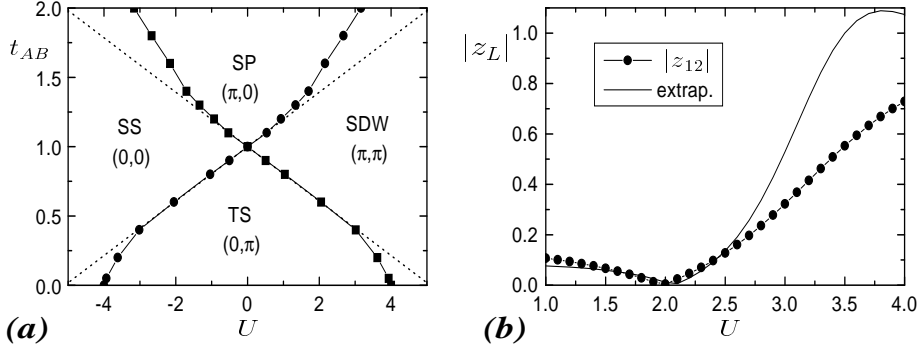,height=2.25in,width=5in}}
%
\caption{ {\it (a)} Quantum phase diagram for $n \rightarrow 1$. The vector 
Berry phase $\vec{\gamma}=(\gamma_c,\gamma_s)$ 
of each thermodynamic phase 
is indicated.
Dotted lines are the phase boundaries according to the CLFT. 
{\it (b)} Localization parameter $|z_L|$ for $t_{AB}=0.6$ as a function of $U$. 
Solid circles are results for $L=12$, and the solid line is a polynomial 
extrapolation in $1/L$ using $L=4,6,8,10$ and 12.}
\label{f1} \end{figure}

For a system of arbitrary length $L$, in contrast to any other observable, 
$\vec{\gamma}$ jumps at the transitions, and the location of this jump in 
parameter space has a very small size dependence. Close to $n=1$, 
the $L \rightarrow \infty$ phase boundaries (extrapolated from $L = 6,8,10$, 
and $12$), are displayed in Fig.~\ref{f1} 
{\it (a)}. We estimate the error in the critical value of $U$ to be less than 1\%.

Dotted lines correspond to results obtained using the CLFT. \cite{jap}
Agreement with the topological transitions is excellent for small $t_{AB}-1$ and $U$, 
in particular, for $0.6\leq t_{AB}\leq 1.2$. For large negative $U$ and $n=1$, the 
dominant correlation functions (CF) at large distances are those for 
the singlet superconductor 
(SS) and a charge density wave (CDW), as in the negative-$U$ Hubbard model. For 
$n\neq 1$, SS CF prevail. For $0<U<4$, $n=1$, and lowering $t_{AB}$, 
the nature of the different thermodynamic phases is the following: a dimerized spin 
(and pseudospin) Peierls (SP) phase, one with dominating spin density wave (SDW) 
CF, and one with identical triplet superconducting (TS) and bond 
SDW (BSDW) CF \cite{top}. We have established the dominant CF 
of this phase using DMRG, since, by symmetry, the 
correlation exponent $K_{\rho }=1$ and all CF decay 
with the same leading power law factor.  
Slightly away from $n=1$, where Umklapp processes in the CLFT become irrelevant, 
$K_{\rho} > 1$ and then, the TS CF are larger than all others at large distances.  
Therefore, in materials with weakly 
coupled chains and $n=1$, fluctuations in the density of each chain should also 
favor the TS CF.

Increasing $U$, there is a Mott transition TS$\rightarrow$SDW which can be 
studied using the quantity $z_{L}=\langle g_{L}|\hat{O}|g_{L}\rangle $, where 
$\hat{O}=\exp (i\frac{2\pi}{L} \sum_{j \sigma} j \, n_{j \sigma})$ and 
$|g_{L}\rangle $ is the GS of the system ($\langle g_{L} |g_{L}\rangle$=1) 
with $L$ sites and $L$ particles \cite{res,z}. Note that $\hat{O}$ shifts the
wave vector of each particle in $2\pi /L$. Then, if $|g_{L}\rangle $ describes a Fermi or 
Luttinger liquid with a well-defined Fermi surface in the thermodynamic 
limit, $\lim_{L \rightarrow \infty} z_{L}=0$ \cite{z}. Instead, for an insulator, 
$|z_{L}|-1$ goes to zero in the same way as the Drude 
weight $D_{c}$ \cite{z}. Furthermore, from a cumulant expansion of $z_L$, it can 
be shown that $\ln |z_{L}|$ is a measure of the fluctuations of the 
polarization (specifically $2(\frac{L}{2\pi})^{2}\ln |z_{L}|=
\langle \sum_{j\sigma} j \, n_{j\sigma} \rangle^{2}
-\langle(\sum_{j\sigma} j \, n_{j\sigma})^{2}\rangle$), which are qualitatively different in an insulator and in a metal. 
The above discussion shows that except 
for some pathological cases, $\lim_{L\rightarrow \infty }|z_{L}|=0$ in a metal, 
while $\lim_{L\rightarrow \infty } |z_{L}|=1$ in an insulator.
In Fig.~\ref{f1}{\it (b)} we show $|z_{L}|$ as a function of $U$ for a case in 
which the topological transition is at $U_{c}=2.051$. For $U<U_{c}$, $|z_{L}|$ 
decreases with $L$, while above $U_{c}$ ($U\sim 3.5-4$), the extrapolated values 
saturate near 1.

We studied the evolution of the phase diagram with doping, using
numerical calculations 
and the extension of the CLFT approach \cite{jap} away from $n=1$. 
For $n=2/3,1/2$, the topological transitions do not completely determine the 
phase diagram, and the calculation of the
appropriate generalization of $\gamma _{c}$ presents some technical problems. 
The jump in $\gamma_{s}$ still very accurately determines the opening of 
$\Delta_s$. For $n=1/2$, the transition to dominating superconducting CF at 
large distances ($K_{\rho }>1$)
is determined using 
numerical data for 8, 12 and 16 sites and finite-size scaling.
While at $n=1$ the computation of $K_{\rho}$ has 
large finite-size effects, for $n=1/2$ we compared $K_{\rho}$ with available 
Bethe-ansatz results for $t_{AB}=1$ \cite{kaw}, with $U=-2$ and $U=-4$ and 
found the difference to be less than 1\%. 

\begin{figure}
\centerline{\psfig{file=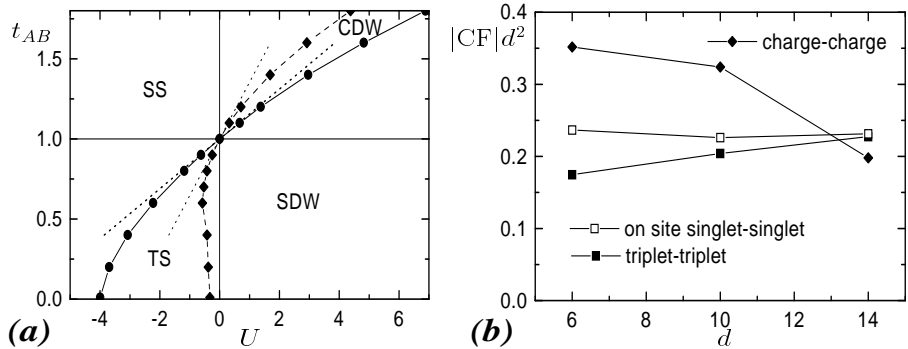,height=2.25in,width=5in}}
%
\caption{{\it (a)} Quantum phase diagram for $n = 1/2$. Circles and diamonds 
denote the closing of $\Delta_s$ and points where $K_\rho=1$, respectively. 
Dotted lines are the phase boundaries according to the CLFT. 
{\it (b)} Modulus of the dominant CF times $d^2$
as functions of distance ($d$) for $t_{AB}=0.6$, $U=-1$.}
\label{f3} \end{figure}

As the system is doped away from 1/2-filling, the SDW advances gradually over the 
TS, and a strong increase in the SS region at the expense of the former SP 
takes place. The numerical phase diagram for $n=1/2$ is shown in Fig. 
\ref{f3}{\it (a)}. Within the CLFT we obtained simple analytical expressions 
$U_{s}=2\sqrt{2}(t_{AB}-1)$ for the transition at $K_{\rho} =1$, and 
$U_{g}=\sqrt{2}(8/\pi+2)(t_{AB}-1)$ for the value of $U$ at which 
$\Delta_s$ closes. As shown in Fig. \ref{f3}{\it (a)}, these expressions  are 
again in excellent agreement with the numerical results, for small $U$ and 
$t_{AB}-1$. However, at intermediate values of the interactions, the deviations 
are much stronger than at 1/2-filling, and the regions in which superconducting 
CF dominate 
are larger than those predicted by the CLFT. The extension of the SS region for 
$t_{AB}>1$ and repulsive $U$ is particularly noticeable.

The nature of each thermodynamic phase, particularly the TS one, was confirmed 
by direct calculation of CF using DMRG. An example 
is shown in Fig. \ref{f3}{\it (b)}. The CF are defined as in Ref. [4].
The slower decay of the TS CF for $t_{AB}=0.6$, and $U=-1$ is consistent with the fact 
that this point lies inside the TS phase in Fig. \ref{f3}{\it (a)}.

In summary,
we determined the phase diagram of a model, expected to describe the 
low-energy physics of some organic superconductors, using recently 
developed geometrical concepts, and other numerical and analytical techniques. 
The phase diagram is very rich. At 1/2-filling ($n=1$), there are all possible 
combinations of open or closed charge and spin gaps, and two different Mott 
transitions take place as a function of the Coulomb on-site interaction $U$. 
There is a wide region in which triplet superconducting CF dominate. This region 
includes positive values of $U$ if $0.57<n<1.43$.

The relevance of the model to (TMTSF)$_{2}$X compounds cannot be
demonstrated until the low-energy reduction is performed. However, it is 
remarkable that for $n\sim 1$ and positive $U$, the sequence of phases 
predicted by the model with decreasing $t_{AB}$ is the same as that
observed with increasing applied pressure \cite{jer}. We expect that $n\sim 1$ 
might be more appropriate than $n\sim 1/2$ if the observed dimerization is 
physically important.


One of us (A.A.A.) thanks L. Arrachea and J. Voit for helpful discussions.
The DMRG numerical calculations were done at the Max-Planck
Institut PKS. A.A.A., K.H. and C.D.B. are
supported by CONICET, Argentina, while G.O. is supported by the US
DOE. K.H. thanks the University of Buenos Aires for hospitality during part of this work. 
This work was supported by PICT 03-00121-02153 of ANPCyT and PIP 4952/96 of CONICET.


\begin{thebibliography}{99}
\bibitem{jer}  D. J\'{e}rome, {\it Science} {\bf 252}, 1509 (1991);
F. Zwick {\it et al.}, {\it Phys. Rev. Lett.} {\bf 79}, 3982 (1997); 
L.P. Gorkov, {\it Sov. Phys. Usp.} {\bf 27}, 809 (1984).

\bibitem{gra}  P.M. Grant {\it et al.}, {\it J. Phys. Colloq. } {\bf 44},
C-3, 847 (1983).

\bibitem{sim}  M.E. Simon, A.A. Aligia, and E.R. Gagliano, {\it Phys. Rev. B}
{\bf 56}, 5637 (1997); references therein.


\bibitem{top}  A.A. Aligia {\it et al.}, cond-mat /9903213; references therein.

\bibitem{res}  R. Resta and S. Sorella, {\it Phys. Rev. Lett.} {\bf 82}, 370
(1999).

\bibitem{z}  A.A. Aligia and G. Ortiz, {\it Phys. Rev. Lett.} {\bf 82}, 2560
(1999).

\bibitem{jap}  J. Japaridze and A. Kampf, {\it Phys. Rev. B} {\bf 59}, 12822
(1999).

\bibitem{kaw}  N. Kawakami and S-K. Yang, {\it Phys. Rev. B} {\bf 44}, 7844
(1991).

\end{thebibliography}
\end{document}